\documentclass[%
reprint,
showpacs,preprintnumbers,
amsmath,amssymb,
longbibliography
]{revtex4-2}

\usepackage{graphicx}
\usepackage{dcolumn}
\usepackage{bm}
\usepackage{colortbl}
\usepackage[unicode=true,colorlinks=true,citecolor=blue,urlcolor=blue]{hyperref}
\usepackage{braket}
\usepackage[normalem]{ulem}

\graphicspath{{img/}}

\newcommand{\e}{\mathrm{e}}

\let\ifr\i
\renewcommand{\i}{{\rm i}}
\renewcommand{\d}{\mathrm d}
\renewcommand{\emph}{\textit}

\renewcommand{\phi}{\varphi}

\usepackage{color}

\begin{document}

\title{Random fine structure and polarized luminescence of triplet excitons\\ in semiconductor nanocrystals}
\author{D.S.~Smirnov\footnote{smirnov@mail.ioffe.ru}, E.L. Ivchenko}
\affiliation{Ioffe Institute, 194021 St. Petersburg, Russia}

\begin{abstract}
  We present a theory of polarized photoluminescence of triplet excitons in semiconductor nanocrystal ensembles with the random fine structure contributed by the electron-hole exchange and carrier-nuclear hyperfine interactions. The interaction parameters are assumed to be normally and isotropically distributed. In particular, the exchange interaction is described by the Gaussian orthogonal ensemble of random matrices. The intensity of luminescence as well as the optical orientation and alignment are calculated as functions of the fine structure splitting parameters and the exciton lifetime. We have also analyzed the suppression of optical alignment and enhancement of optical orientation in an external longitudinal magnetic field.
\end{abstract}

\maketitle

\section{Introduction}
In direct band-gap semiconductors with simple conduction and valence bands, the ground exciton states form a quartet of sublevels. In crystals or nanocrystals of cubic symmetry, due to the electron-hole exchange interaction the quartet undergoes a singlet-triplet splitting \cite{Goupalov3,Goupalov1}. In anisotropic crystals the exciton triplet also splits into two or three terms. For example, in uniaxial GaSe crystals the free or bound exciton triplet is split into optically active doublet and singlet \cite{Theory,Experiment,GaSe-bound}. The triplet splitting occurs in cubic-based semiconductor quantum dots and nanocrystals because of the anisotropic shape of quantum-confining potential \cite{Goupalov1,Goupalov2}. The nanocrystal symmetry imposes the optical selection rules for the split excitonic terms as it takes place in GaSe crystals.

We consider an ensemble of cubic-based nanocrystals with strong singlet-triplet splitting and random confining potential leading to the triplet exchange Hamiltonian described by a manifold of random matrices. The field related to random matrices play an important role in areas of pure mathematics and physical applications \cite{Dyson,PRL1991,Anderson}. In addition to the exchange interaction, we consider random Overhauser fields caused by exciton-nuclear hyperfine interaction. Then we analyze the consequences of these two kinds of interaction for the polarized exciton photoluminescence (PL) in the presence of an external magnetic field.

\section{Formalism}

We consider a triplet of bright exciton states and ignore the dark singlet state assuming the splitting between them to be large enough. Then the exciton Hamiltonian can be generally written in terms of the exciton angular momentum operator $\bm L$ ($L=1$) with the components $L_{\alpha}~(\alpha= x,y,z)$ being 3$\times$3 matrices. Its most general form is
\begin{equation}
  \label{eq:H}
  \mathcal H=\hbar\sum_{\alpha}\Omega_\alpha L_\alpha+\hbar\sum_{i}\delta_iD_i,
\end{equation}
where $i=yz$, $xz$, $xy$, $x^2-y^2$, $z^2$, the corresponding operators read
\begin{eqnarray} \label{Dij}
&&  D_{yz}=L_yL_z+L_zL_y \:,\: D_{xz}=L_xL_z+L_zL_x\:,   \\ &&
D_{xy}=L_xL_y+L_yL_x \:,\:  D_{x^2-y^2}=L_x^2-L_y^2 \:, \nonumber\\ &&
D_{z^2}=\frac{1}{\sqrt{3}}\left(2L_z^2-L_x^2-L_y^2\right) \:, \nonumber
\end{eqnarray}
and $\Omega_\alpha$, $\delta_i$ are real coefficients in units of inverse time. Note that $\bm L$ is odd under time reversal, therefore $\bm \Omega$ describes the effect of external or effective magnetic fields acting on the exciton, while $\delta_i$ describes exchange splitting of the exciton sublevels which is even under time reversal. This form of the Hamiltonian is invariant and can be written in any basis of the exciton states. In the following we use the notation ${\cal H}_{\Omega}$ and ${\cal H}_{\delta}$ for the first and second sums in the right-hand side of Eq.~\eqref{eq:H}.

In an ensemble of nanocrystals, the excitonic parameters $\Omega_\alpha$ and $\delta_i$ can be different for different nanocrystals, and one needs to average the observables over the distribution of these parameters. In this section, we focus on one particular nanocrystal. The averaging is discussed in detail in the next section.

The exciton dynamics can be described by the exciton density matrix $\rho_{mm'}$, where the index $m$ enumerates the eigenstates of the Hamiltonian~\eqref{eq:H}. It satisfies the master equation
\begin{equation}
  \label{eq:drho}
  \frac{\d\rho}{\d t}=-\frac{\i}{\hbar}\left[\mathcal H,\rho\right]-\frac{\rho}{\tau}+g\:,
\end{equation}
where $\tau$ is the exciton radiative time, assumed to be the same for all three exciton states and $g$ is the exciton generation matrix. For the resonant exciton generation, the latter has the form
\begin{equation}
  \label{g}
  g_{mm'} \propto \sum_{\alpha\beta} M_{m,\alpha}M^*_{m',\beta} d^0_{\alpha\beta}\:,
\end{equation}
where $M_{m,\alpha}$ is the matrix element of optical excitation of the exciton $m$ by the light with polarization $\alpha$, and $d^0_{\alpha\beta}$ is the polarization matrix of the exciting light. In the particular basis of excitonic states $|m \rangle$ optically active in the polarizations $\alpha = x,y,z$, respectively, and for the light propagating along the $z$ axis, we can replace $m,m'$ by $x,y,z$ and sum in Eq.~\eqref{g} over $\alpha,\beta=x,y$ only. In this basis, one has $M_{m,\alpha}\propto \delta_{m,\alpha}$. The selection rules for arbitrary exciton basis states are obtained from these relations by a corresponding unitary transformation.

Equation~\eqref{eq:drho} takes a particularly simple form in the steady state and in the basis of the exciton eigenstates:
\begin{equation}
  \label{tau}
  \left[\frac{1}{\tau} + {\rm i} (\omega_m - \omega_{m'})\right] \rho_{mm'} = g_{mm'}\:,
\end{equation}
where $\omega_m$ are the exciton eigenfrequencies. From this equation one can readily find all the density matrix elements $\rho_{mm'}$. They determine the secondary radiation density (or polarization) matrix $d_{\alpha\beta}$ as
\begin{equation}
  \label{d}
  d_{\alpha \beta}\propto \sum\limits_{mm'} M^*_{m,\alpha}M_{m',\beta} \rho_{mm'}\:.
\end{equation}
Making use of Eqs.~\eqref{g} and~\eqref{tau} it gives
\begin{equation}
  \label{tau2}
  d_{\alpha \beta}\propto \sum\limits_{\substack{mm'\\\alpha'\beta'}} \frac{M^*_{m,\alpha}M_{m',\beta}M_{m,\alpha'}M^*_{m',\beta'} }{\tau^{-1} + {\rm i} (\omega_m - \omega_{m'})}\ d^0_{\alpha' \beta'}.
\end{equation}
This is a general expression that relates the polarization matrices of the secondary emission and the incident light.

For example, in the limit of small exciton splittings or short exciton lifetime, $(\omega_m-\omega_{m'})\tau\to0$, we obtain
\begin{equation}
  d_{\alpha \beta}\propto\sum\limits_{mm'} M^*_{m,\alpha}M_{m',\beta} \sum\limits_{\alpha' \beta'} M_{m,\alpha'}M^*_{m',\beta'}\ d^0_{\alpha' \beta'} = d^0_{\alpha \beta}\:
\end{equation}
so the polarization matrices of the incident and emitted light are proportional to each other. In particular, all their Stokes parameters coincide, as expected.

In the opposite limit of the large splitting and the long exciton lifetime, $(\omega_m-\omega_{m'})\tau\to\infty$ for $m\neq m'$, we obtain
\begin{equation}
  \label{eq:splitted}
  d_{\alpha \beta}\propto \sum\limits_{\substack{m\\\alpha'\beta'}} M^*_{m,\alpha}M_{m,\beta}M_{m,\alpha'}M^*_{m,\beta'}d^0_{\alpha' \beta'}.
\end{equation}
This expression neglects the interference of light emitted by different exciton states and can be obtained as a result of independent excitation of states $m$ by the polarized light.

In the next section we analyze the transition between these two regimes for particular cases of the fine structure of exciton ensemble.

\section{Exciton random fine structure}

\subsection{Random exchange splittings}
\label{sec:exchange}

First, we consider the case, when no (effective) magnetic fields act on excitons, and set $\Omega_\alpha=0$. The remaining parameters $\delta_i$ describe time-reversal-invariant exchange splittings. We assume that they are caused by various weak, random and uncorrelated perturbations such as strain in the glass matrix and nanocrystal shape distortions. We further assume that the ensemble of nanocrystals is isotropic on average. The operators $D_i$ in Eqs.~\eqref{eq:H}, \eqref{Dij} transform as the real spherical harmonics of the angular momentum $2$ or the atomic $d$-orbitals. Therefore, the corresponding coefficients are independent, identically and normally distributed. Thus, their probability density function has the form
\begin{equation}
  \mathcal P(\delta_i)=\frac{1}{\sqrt{\pi}\delta}\exp(-\delta_i^2/\delta^2),
\end{equation}
where $\delta$ is the typical exchange splitting of the exciton states (in units of inverse time).

It is convenient to write the real Hamiltonian ${\cal H}_{\delta}$ in the basis of $x$, $y$, and $z$-polarized exciton states. In this basis, one has \cite{Varshalovich,OO_book} 
\begin{eqnarray}
&&L_x= \left[ \begin{array}{ccc}
0&0 &0\\
0&0 &$- {\rm i}$\\
0&${\rm i}$ &0
\end{array}
   \right]\:,\:
   \quad
   L_y= \left[ \begin{array}{ccc}
0&0 &${\rm i}$\\
0&0 &0\\
$- {\rm i}$&0 &0
\end{array}
\right]\:,\\&& \hspace{2 cm}L_z= \left[ \begin{array}{ccc}
0&$- {\rm i}$ &0\\
${\rm i}$&0 &0\\
0&0 &0
\end{array}
\right]\:. \nonumber
\end{eqnarray}

The eigenstates ${\bm \psi}^{(m)}$ of ${\cal H}_{\delta}$ satisfy the Schr\"odinger equation
\begin{equation} \label{Sch}
{\cal H}_{\delta}{\bm \psi}^{(m)} = E_m {\bm \psi}^{(m)}\:,
\end{equation}
where $E_m = \hbar \omega_m$.
We denote their components as $\psi^{(m)}_\alpha$, $\alpha=x,y,z$. From Eq.~\eqref{Sch} we obtain
\begin{equation} \label{xyz}
\psi^{(m)}_x = u^{(m)}_x \psi^{(m)}_z\:,\: \psi^{(m)}_y = u^{(m)}_y \psi^{(m)}_z\:,
\end{equation}
where
\begin{eqnarray} \label{C1C2}
&& u^{(m)}_x = \frac{ (E_m - H_{yy}) H_{xz} + H_{xy} H_{yz} }{(H_{xx} - E_m)(H_{yy} - E_m) - H_{xy}^2} \:, \\
&& u^{(m)}_y = \frac{ (E_m - H_{xx}) H_{yz} + H_{xy} H_{xz}}{(H_{xx} - E_m)(H_{yy} - E_m) - H_{xy}^2} \:. \nonumber
\end{eqnarray}

\begin{figure}
  \includegraphics[width=0.5\linewidth]{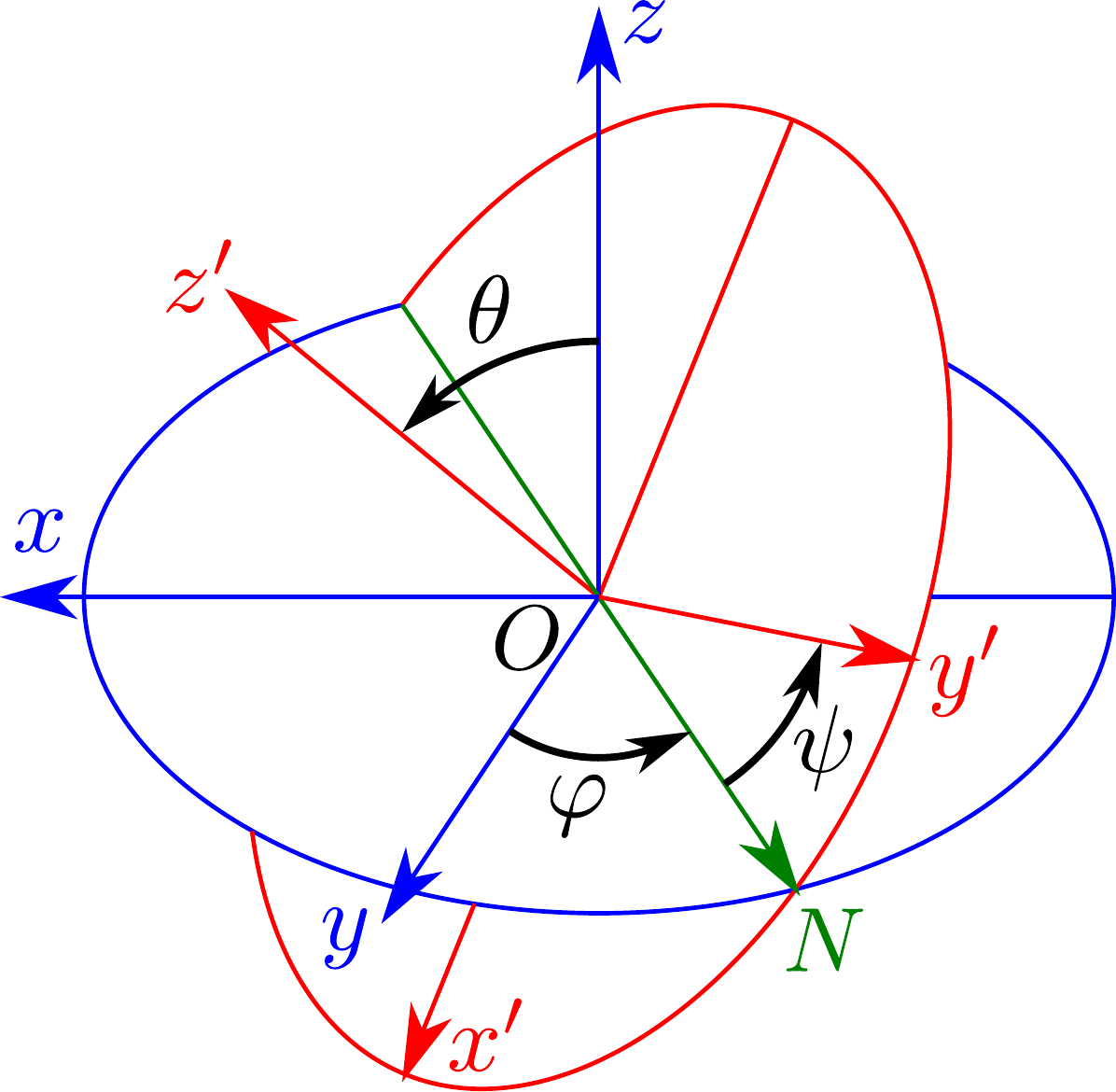}
  \caption{
    \label{fig:Euler}
    Definition of the Euler angles. First, the initial $(xyz)$ system is rotated about the $z$ axis by $\varphi$. The new $y$ axis makes the angle $\varphi$ with the initial $y$ axis. Then the new system is rotated about its $y$ axis (denoted $ON$) by $\theta$. As a result, the new $z$ axis makes the angle $\theta$ with the initial $z$ axis. Finally, the system is rotated about its current $z$ axis by the angle $\psi$. The final coordinate system is denoted $(x'y'z')$.
  }
\end{figure}

The eigenstates ${\bm \psi}^{(m)}$ are real, orthogonal, and invariant under time reversal. This means that they represent the linearly polarized exciton states along some Cartesian axes $x'$, $y'$, and $z'$ related to $x$, $y$, and $z$ by three Euler angles $\varphi, \theta$, and $\psi$ defined in Fig.~\ref{fig:Euler} according to Ref.~\onlinecite{LL3}. The connection of these angles to the Hamiltonian ${\cal H}_{\delta}$ is established in Appendix~\ref{app:wf}. 
In terms of $\varphi, \theta$, and $\psi$, the eigenstates have the form
\begin{eqnarray} \label{Cnew}
&&{\bm \psi}^{(x')}= \left[ \begin{array}{c} \cos{\theta} \cos{\varphi} \cos{\psi} - \sin{\varphi} \sin{\psi} \\ \cos{\theta} \sin{\varphi} \cos{\psi} + \cos{\varphi} \sin{\psi} \\ - \sin{\theta}\cos{\psi} \end{array} \right]\:, \\&&{\bm \psi}^{(y')}= \left[ \begin{array}{c}  - \cos{\theta} \cos{\varphi} \sin{\psi} - \sin{\varphi}\cos{\psi} \\- \cos{\theta} \sin{\varphi} \sin{\psi} + \cos{\varphi}\cos{\psi}\\ \sin{\theta} \sin{\psi} \end{array} \right]\:,\nonumber\\&& {\bm \psi}^{(z')}= \left[ \begin{array}{c} \sin{\theta} \cos{\varphi} \\ \sin{\theta} \sin{\varphi}\\ \cos{\theta} \end{array} \right]  \:. \nonumber
\end{eqnarray}

The matrix elements of the optical transitions $M_{m,\alpha}$ with $m=x'$, $y'$, and $z'$ are related to ${\bm \psi}^{(m)}$ by
\begin{equation} \label{eq:M}
M_{m,\alpha} \propto \psi^{(m)*}_{\alpha} \:.
\end{equation}
Since here the values of $\psi^{(m)}_{\alpha}$ are real the complex conjugation can be omitted.

Substituting Eq.~\eqref{eq:M} into Eq.~\eqref{tau2}, averaging over the Euler angles and omitting the constant factor we obtain for the PL intensity
\begin{equation}
  \label{eq:int_delta}
I \equiv  d_{xx} + d_{yy} = \frac15 (4+R)\  (d^0_{xx} + d^0_{yy})\:,
\end{equation}
where 
\begin{equation} \label{eq:R}
\hspace{-0.3 cm}  R=\frac{1}{3}\left(  \frac{1}{1 + \omega^2_{x'y'} \tau^2}+ \frac{1}{1 + \omega^2_{x'z'} \tau^2} + \frac{1}{1 + \omega^2_{y'z'} \tau^2} \right)
\end{equation}
with $\omega_{mm'}=\omega_{m}-\omega_{m'}$. In the same way, we obtain for the difference of intensities in linear polarizations
\begin{equation}
  \label{eq:dxy_delta}
  d_{xx} - d_{yy} = \frac{2+3R}{5}\ (d^0_{xx} - d^0_{yy}).
\end{equation}

The ensemble of random real matrices described by five independent and identically distributed parameters $\delta_i$ in Eq.~\eqref{eq:H} is a particular case of Gaussian orthogonal ensemble~\cite{Dyson,Mehta,Anderson}. The joint probability distribution of its eigenfrequencies reads~\cite{Selberg,Dyson,PRL1991,Mehta}
\begin{multline}
  \label{eq:P}
  \mathcal P(\omega_{x'},\omega_{y'},\omega_{z'})\\=\frac{|\omega_{x'y'}\omega_{x'z'}\omega_{y'z'}|}{6\sqrt{2}\pi\delta^6}\exp\left(-\frac{\omega_{x'}^2+\omega_{y'}^2+\omega_{z'}^2}{2\delta^2}\right).
\end{multline}
This distribution function allows us to calculate the average of $R$ in Eq.~\eqref{eq:R} and thus to find the intensities of linearly polarized PL components.

\begin{figure*}
  \includegraphics[width=0.49\linewidth]{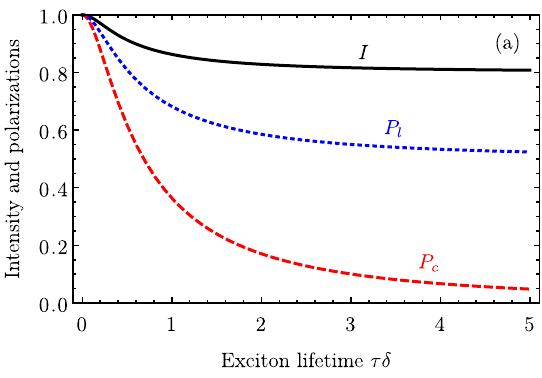}~
  \includegraphics[width=0.49\linewidth]{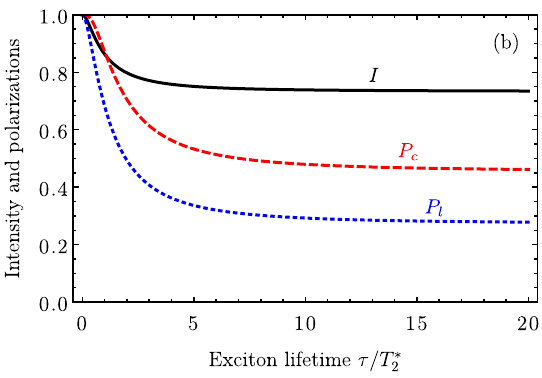}
  \caption{
    \label{fig:tau}
    Variation of the PL intensity (black solid lines) and degrees of the linear (blue dotted lines) and circular (red dashed lines) polarizations with the exciton lifetime $\tau$. (a) Effect of the exchange interaction calculated after Eqs.~\eqref{eq:int_delta}, \eqref{eq:Pl_delta}, and \eqref{eq:Pc_delta}. (b) Effect of the hyperfine interaction calculated after Eqs.~\eqref{eq:int_hf} and \eqref{eq:P_hf}.
  }
\end{figure*}

The black solid line in Fig.~\ref{fig:tau}(a) shows the dependence of the total PL intensity $I$ on the product $\tau \delta$ which describes the interplay between the exciton recombination and exchange interaction. The intensity slightly decreases by $20$\% with the increasing lifetime, as it readily follows from Eq.~\eqref{eq:int_delta}. This is a surprising result, because for conventional GaAs-based quantum dots (with a different exciton fine structure), the intensity is determined only by the ratio of radiative and nonradiative lifetimes and is independent of the strength of exchange interaction. Thus, the variation of $I$ with $\tau$ is a signature of the triplet (instead of doublet in GaAs-based quantum dots) character of bright excitons under consideration.

We remind that the optical orientation and alignment are two important examples of the selective excitation of excitonic sublevels by polarized light~\cite{OptOr}. In the effect of optical orientation the circularly polarized light causes a preferential orientation of the exciton spin (or angular momentum) which manifests itself in an appearance of the PL circular polarization. The optical alignment means a preferential orientation of the exciton oscillating dipole moment in a certain direction and an appearance of the PL linear polarization under linearly polarized photoexcitation. The PL linear polarization is described by the Stokes parameter 
\begin{equation}
  P_l = \frac{d_{xx} - d_{yy}}{d_{xx} + d_{yy}}\:.
\end{equation}
According to Eqs.~\eqref{eq:int_delta} and~\eqref{eq:dxy_delta}, under resonant linearly polarized photoexcitation the degree of exciton alignment is given by
\begin{equation}
  \label{eq:Pl_delta}
  P_l=\frac{2+3R}{4+R}
\end{equation}
where the averaging of $R$ with the distribution~\eqref{eq:P} is implicitly assumed. The function $P_l(\tau \delta)$ is shown in Fig.~\ref{fig:tau}(a) by the blue dotted line. One can see the depolarization from $100\%$ to $50\%$ with the increasing value of $\tau \delta$.

The optical orientation of excitons can be calculated in the basis of circularly polarized components $\sigma_+$ and $\sigma_-$. The result reads 
\begin{equation}
  d_{\sigma_+ \sigma_+} - d_{\sigma_- \sigma_-} = R\ (d^0_{\sigma_+ \sigma_+} - d^0_{\sigma_- \sigma_-})\:,
\end{equation}
from which we obtain the degree of circular polarization
\begin{equation}
  \label{eq:Pc_delta}
  P_c= \frac{d_{\sigma_+ \sigma_+} - d_{\sigma_- \sigma_-}}{d_{xx} + d_{yy}} = \frac{5R}{4+R}.
\end{equation}
Variation of $P_c$ with $\tau \delta$ is illustrated in Fig.~\ref{fig:tau}(a) by the red dashed line. For short exciton lifetimes, the polarization is completely preserved, $P_c=1$, but with increasing $\tau \delta$ it tends to zero. The reason is that each of the exciton eigenstates split by the exchange interaction emits linearly polarized light.

\subsection{Hyperfine interaction}
\label{sec:hf}

The strength of exchange interaction within the exciton crucially depends on the electron--hole overlap in real and reciprocal spaces. In this regard, quantum dots and nanocrystals can be classified into one of four types based on the character of this overlap \cite{SmirnIvch1,SmirnIvch2}: d-$r$--d-$k$, ind-$r$--d-$k$, d-$r$--ind-$k$, and ind-$r$--ind-$k$, where ``d'' and ``ind'' mean direct and indirect, and $r, k$ mean the ${\bm r}$ and ${\bm k}$ spaces, respectively. Nano-engineering provides control over the electron--hole overlap and reduction of the exchange interaction in the nanostructures different from the d-$r$--d-$k$ type, see, e.g., Ref.~\cite{core-shell}. In this subsection, we consider nanocrystal arrays where the exchange splitting parameters $\delta_i$ in Eq.~\eqref{eq:H} are smaller than the hyperfine interaction, while the bright--dark splitting remains larger and the dark exciton singlet state can still be neglected.

Thus, we set $\delta_i=0$ in Eq.~\eqref{eq:H} but keep the Overhauser frequency $\Omega_\alpha$ replacing the Hamiltonian ${\cal H}$ by ${\cal H}_{\Omega}$. Since each charge carrier in a nanocrystal interacts with a large number of lattice nuclei, the distribution of $\bm\Omega$ is Gaussian. Following Refs.~\cite{PolarizedNuclei,bookGlazov,ZenoExp} we write it as
\begin{equation}
  \label{eq:F}
  \mathcal F(\bm\Omega)=\left(\sqrt{\frac{2}{\pi}}T_2^*\right)^3 \e^{-2 \left( \Omega T_2^* \right) ^2}\:,
\end{equation}
where $T_2^*$ is the exciton spin dephasing time. The nuclear spin dynamics is much slower than the exciton recombination time, hence $\bm\Omega$ can be assumed static~\cite{Merkulov,bookGlazov}. To stress the difference from the Gaussian orthogonal ensemble, the random distribution of matrices ${\cal H}_{\Omega}$ can be coined as the Gaussian pseudomagnetic ensemble.

The exciton eigenstates $m$ in this case are simply the states with the angular momentum projection $m \equiv L_{z'} = 0, \pm 1$ onto the direction $z' \parallel \bm\Omega$. By using Wigner D-functions~\cite{Varshalovich} they can be obtained from the states with the angular momentum component $L_z=0,\pm1$ along the $z$ direction of the optical axis. The energies are just equal to $m \hbar\Omega$. Then, following the lines of the previous subsection, we obtain after averaging over the directions of~$\bm\Omega$:
\begin{subequations}
   \begin{eqnarray}
&&   \label{eq:int_hf}
  d_{xx} + d_{yy} = \frac{11+2Q(\tau)+2Q(2\tau)}{15}  (d^0_{xx} + d^0_{yy}), \hspace{2 mm}\\
    && \label{eq:dxy_hf}
  d_{xx} - d_{yy} = \frac{1+2Q(\tau)+2Q(2\tau)}{5} (d^0_{xx} - d^0_{yy}),\\
    && \label{eq:dpm_hf}
  d_{\sigma_+ \sigma_+} - d_{\sigma_- \sigma_-} = \frac{1+2Q(\tau)}{3} (d^0_{\sigma_+ \sigma_+} - d^0_{\sigma_- \sigma_-}),
  \end{eqnarray}
\end{subequations}
where
\begin{equation}
  \label{eq:R1}
Q(\tau)=\frac{1}{1 + (\Omega\tau)^2}\;.
\end{equation}
Similarly to $R(\tau)$ in Eq.~\eqref{eq:R}, the function $Q(\tau)$ as well as $Q(2\tau)$ decays from $1$ to $0$ with an increase of $\tau$ from zero to infinity; $Q(\tau)$ and $Q(2\tau)$ should be also averaged over the absolute values of $\Omega$ with the distribution function~\eqref{eq:F}. The averaging should be performed as
  \begin{equation}
    \label{Lz2}
    \langle Q(\tau) \rangle = \int\limits_0^{\infty}   \frac{ 4 \pi \Omega^2 \mathcal F(\Omega)}{1 + \Omega^2 \tau^2} \d\Omega\:.
\end{equation}

\begin{figure*}
  \includegraphics[width=0.49\linewidth]{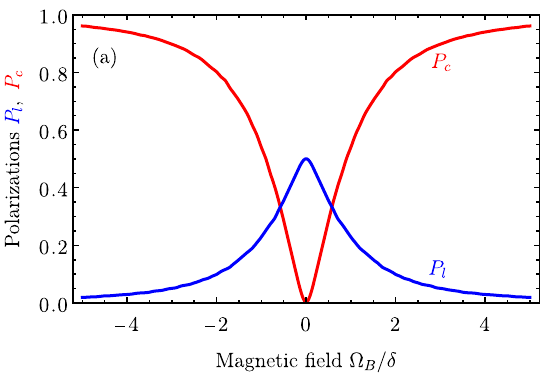}
  \includegraphics[width=0.49\linewidth]{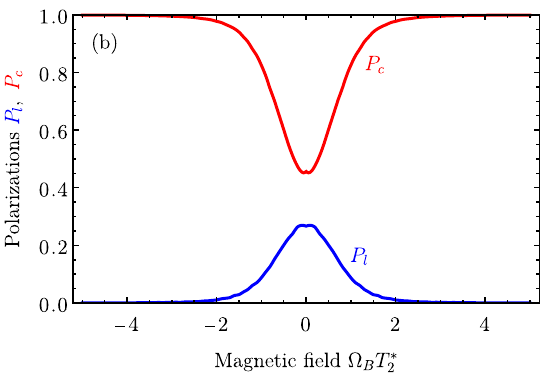}
  \caption{
    \label{fig:Bz2}
    Dependencies of the optical orientation (red lines) and optical alignment (blue lines) on the external longitudinal magnetic field in cases of the random exchange (a) and hyperfine (b) interaction. 
  }
\end{figure*}

Equation~\eqref{eq:int_hf} determines the PL intensity, while Eqs.~\eqref{eq:dxy_hf} and~\eqref{eq:dpm_hf} give the degrees of linear and circular polarizations
\begin{subequations}
  \label{eq:P_hf}
\begin{eqnarray}
&&  \label{eq:Pl_hf}
  P_l=\frac{3+6Q(\tau)+6Q(2\tau)}{11+2Q(\tau)+2Q(2\tau)}\:,\\
&&  \label{eq:Pc_hf}
  P_c=\frac{5+10Q(\tau)+6Q(2\tau)}{11+2Q(\tau)+2Q(2\tau)}\:.
\end{eqnarray}
\end{subequations}
These Stokes parameters are shown in Fig.~\ref{fig:tau}(b) as functions of the ratio of the exciton lifetime and spin dephasing time. Comparing Figs.~\ref{fig:tau}(a) and~\ref{fig:tau}(b) we see remarkable differences. First, the intensity in Fig.~\ref{fig:tau}(b) decreases asymptotically not by $20$\% but by more than $26$\%. Second, the optical orientation is higher than the optical alignment. Moreover, the optical orientation does not vanish but saturates at $5/11\approx 0.45$ for long exciton lifetimes. This result is similar to the electron spin dephasing in the quantum dots where the spin polarization reduces by a factor of 3~\cite{Merkulov}. The same happens here with the exciton spin polarization, but for the circular PL polarization degree of triplet excitons, the depolarization is weaker because $P_c$ is not determined exclusively by the spin polarization. Third, the optical alignment does not decay to zero in the limit of large $\tau/T_2^*$, but saturates at $3/11\approx0.27$ despite the formation of circular exciton eigenstates. This happens because the exciton spin quantization axis $z'$ deviates from the $z$ axis, and the exciton circular eigenstates can emit partially linearly polarized light in the $z$ direction.

\section{Role of magnetic field}

In this section, we consider the limit of large exciton splittings as compared with the exciton lifetime and study the role of the external longitudinal magnetic field applied in the Faraday geometry. Generally, the secondary emission polarization matrix is given by Eq.~\eqref{eq:splitted} in this limit.

\subsection{Gaussian orthogonal ensemble}

We start the analysis by assuming the exchange interaction to be stronger than the hyperfine interaction similarly to Sec.~\ref{sec:exchange}. The corresponding exciton Hamiltonian has the form
\begin{equation}
  \mathcal H=\hbar\sum_{i}\delta_iD_i+\hbar\Omega_B L_z,
\end{equation}
where $\Omega_B$ is the exciton Larmor spin precession frequency in the external magnetic field ${\bm B} \parallel z$. It is given by $\Omega_B=g\mu_BB/\hbar$, where $g$ is the exciton $g$-factor and $\mu_B$ is the Bohr magneton. We find the exciton eigenfunctions and eigenfrequencies numerically and show the result of the calculation of the Stokes parameters in Fig.~\ref{fig:Bz2}(a).

One can see that the optical alignment is suppressed and the optical orientation is recovered by the longitudinal magnetic field similar to the excitons in GaAs-based quantum dots~\cite{dzhioev1997,ivchenko05a}. This is caused by the formation of the circularly polarized exciton states with $L_z=0,\pm1$ at $\Omega_B\gg\delta$. Note that, in agreement with Eq.~\eqref{eq:Pl_delta}, the linear polarization degree in zero magnetic field equals $50$\%.

The exciton PL intensity in the $z$ direction also changes with the increasing magnetic field, it increases by $20$\%, in contrast to the solid line in Fig.~\ref{fig:tau}(a).

\subsection{Gaussian pseudomagnetic ensemble}

Now we turn to the interplay between the external magnetic field and the random Overhauser field neglecting the exchange splitting between the bright exciton sublevels. It is described by the Hamiltonian
\begin{equation}
  \mathcal H=\hbar\bm\Omega\bm L+\hbar\Omega_BL_z,
\end{equation}
where the Overhauser frequency $\bm\Omega$ is distributed according to Eq.~\eqref{eq:F}.

The calculated dependencies of the optical orientation $P_c$ and optical alignment $P_l$ on the magnetic field are shown in Fig.~\ref{fig:Bz2}(b). In contrast to Fig.~\ref{fig:Bz2}(a), at zero magnetic field they are both nonzero and equal to $5/11$ and $3/11$, respectively, in agreement with Eqs.~\eqref{eq:Pc_hf} and~\eqref{eq:Pl_hf}. Application of strong enough magnetic field, $\Omega_B\gg1/T_2^*$, again suppresses the optical alignment to zero and recovers the optical orientation to $100$\%. At the same time, the PL intensity slightly increases (not shown).

\section{Outlook and conclusion}

As an outlook, we note that the theory of random matrices can be applied for analysis of the triplet exciton polarized photoluminescence in perovskite nanocrystal ensembles, because of the simple structure of their conduction and valence bands in the corresponding bulk materials~\cite{Nestoklon,Yakovlev}. Noteworthy, in addition to the random and fixed fine structure, there are ensembles with different types of randomness which fill a gap between the two alternatives. For illustration, we refer to the work~\cite{Dzioev} on InAlAs/AlGaAs quantum dots where the value of $\delta_{x^2 - y^2}$ in Eq.~\eqref{eq:H} is random whereas the average of $\delta_{xy}$ is nonzero.

To conclude, we have calculated the intensity and Stokes parameters of resonantly excited photoluminescence of zero-dimensional triplet excitons in an ensemble of nanocrystals with the random fine structure. We have taken into account the interplay between exciton lifetime, external magnetic field, hyperfine interaction between the charge carriers and lattice nuclei and the electron-hole exchange interaction. The latter is described by the Gaussian orthogonal ensemble of Hamiltonian matrices.

\section*{Acknowledgments}

We thank D.R.~Yakovlev, T.S. Shamirzaev and I.A. Yugova for helpful discussions. D.S.S. acknowledges support of Foundation for the Advancement of Theoretical Physics and Mathematics ``BASIS''. The work is also supported by the Russian Science Foundation Grant 25-72-10031.

\appendix

\section{Mapping of the real Hamiltonian ${\cal H}_{\delta}$ on the Euler angles}
\label{app:wf}
It should be kept in mind that the eigenstates of the Hamiltonian ${\cal H}_{\delta}$ are specified only up to a sign and that the polarization matrix $d_{\alpha \beta}$ in Eq.~\eqref{tau2} is independent of this sign. To be specific, we set $\psi^{(m)}_z > 0$, so that the columns ${\bm \psi}^{(m)}$ can be written as
\begin{equation} \label{psi+}
{\bm \psi}^{(m)} = \frac{1}{ \sqrt{ 1 + \left( u_x^{(m)} \right)^2 + \left( u_y^{(m)} \right)^2 } } \left[ \begin{array}{c} u^{(m)}_x \\  u^{(m)}_y \\ 1 \end{array} \right]\:.
\end{equation} 
The three-component columns can be usefully considered as three-dimensional vectors. We choose the axis $z'$ to correspond to the largest energy $E_m$, i.e., $E_{z'} = \max{(E_m)}$. The order {of functions ${\bm \psi}^{(x')}$ and ${\bm \psi}^{(y')}$ is determined by a positive value of the scalar triple product $\left( {\bm \psi}^{(x')} \times {\bm \psi}^{(y')}\right) \cdot {\bm \psi}^{(z')}$.

Comparing the columns ${\bm \psi}^{(z')}$ in Eqs.~\eqref{Cnew} and~\eqref{psi+} we find
\begin{equation} \label{cos}
\cos{\theta} = \left[1 + \left(u_{x}^{(z')} \right)^2 + \left(u_{y}^{(z')} \right)^2 \right]^{-1/2}
\end{equation}
and
\begin{equation} \label{psi}
\sin{\varphi} = \tan{\theta}\ u_{x}^{(z')}\:,\\  \cos{\varphi} =\tan{\theta}\ u_{y}^{(z')}\:. \end{equation}
Comparing the columns ${\bm \psi}^{(x')}, {\bm \psi}^{(y')}$ in Eqs.~\eqref{Cnew} and~\eqref{psi+} we obtain
\begin{eqnarray} \label{psi}
&&\sin{\psi} =  \frac{1}{\sin{\theta}}\left[1 + \left(u_{x}^{(y')} \right)^2 + \left(u_{y}^{(y')} \right)^2 \right]^{-1/2}\:,\\ &&  \cos{\psi} = - \frac{1}{\sin{\theta}}\left[1 + \left(u_{x}^{(x')} \right)^2 + \left(u_{y}^{(x')} \right)^2 \right]^{-1/2}\:. \nonumber
\end{eqnarray}

The obtained angles satisfy the conditions $\cos{\theta} > 0$, $\sin{\psi} > 0$, and $\cos{\psi} < 0$. Other regions for $\theta$ and $\psi$ {can be reached by changing the sign of the columns~\eqref{psi+}. This means that the polarization tensor $d_{\alpha \beta}$ given by Eq.~\eqref{tau2} can be averaged over all values of $\varphi$, $\theta$, and $\psi$.

\renewcommand{\i}{\ifr}

\end{document}